\documentclass[showpacs,aps,prl,reprint,superscriptaddress,preprintnumbers,nolinenumbers]{revtex4-2}

\makeatletter
\newcommand{\fmarki}{*}
\newcommand{\fmarkii}{\ensuremath{\dagger}}
\def\@fnsymbol#1{{\ifcase#1\or \fmarki\or \fmarkii\or \fmarkiii\or \fmarkiv\or \fmarkv\or \fmarkvi\or \fmarkvii\or \fmarkviii\or \fmarkix \else\@ctrerr\fi}}
\makeatother
\renewcommand{\fmarkii}{*}

\usepackage{amsmath}
\usepackage{amssymb}
\usepackage{graphicx}
\usepackage{dcolumn}
\usepackage[dvipsnames]{xcolor}
\usepackage{amsthm,amssymb}
\usepackage{mathrsfs}
\usepackage{color}
\usepackage{endnotes}
\usepackage{bm}
\usepackage{epsfig}
\usepackage{array}
\usepackage{ulem}

\newcommand{\ie}{{\it i.e.\ }}
\newcommand{\eg}{{\it e.g.\ }}

\begin{document}
\title{Tunable non-Lifshitz-Kosevich temperature dependence of \\ Shubnikov-de Haas oscillation amplitudes in SmSb}

\author{Wei~Zhang}
\affiliation{Department of Physics, The Chinese University of Hong Kong, Shatin, Hong Kong, China} 
\author{C.~N.~Kuo}
\author{S.~T.~Kuo}
\affiliation{Department of Physics, National Cheng Kung University, Tainan 70101, Taiwan}
\affiliation{Taiwan Consortium of Emergent Crystalline Materials, Ministry of Science and Technology, Taipei 10601, Taiwan}
\author{Chun~Wa~So}
\affiliation{Department of Physics, City University of Hong Kong, Kowloon, Hong Kong, China}
\author{Jianyu Xie}
\affiliation{Department of Physics, The Chinese University of Hong Kong, Shatin, Hong Kong, China} 
\author{Kwing~To~Lai}
\affiliation{Department of Physics, The Chinese University of Hong Kong, Shatin, Hong Kong, China} 
\affiliation{Shenzhen Research Institute, The Chinese University of Hong Kong, Shatin, Hong Kong, China}
\author{Wing~Chi~Yu}
\affiliation{Department of Physics, City University of Hong Kong, Kowloon, Hong Kong, China}
\author{C.~S.~Lue}
\affiliation{Department of Physics, National Cheng Kung University, Tainan 70101, Taiwan}
\affiliation{Taiwan Consortium of Emergent Crystalline Materials, Ministry of Science and Technology, Taipei 10601, Taiwan}
\author{Hoi Chun Po}
\email[]{hcpo@ust.hk}
\affiliation{Department of Physics, The Hong Kong University of Science and Technology, Clear Water Bay, Kowloon, Hong Kong, China}
\author{Swee~K.~Goh}
\email[]{skgoh@cuhk.edu.hk}
\affiliation{Department of Physics, The Chinese University of Hong Kong, Shatin, Hong Kong, China}

\date{\today}

\begin{abstract} 
The Lifshitz-Kosevich (LK) theory is the pillar of magnetic quantum oscillations, which have been extensively applied to characterize a wide range of metallic states. In this study, we focus on the Shubnikov-de Haas (SdH) effect observed in SmSb, a rare-earth monopnictide. We observed a significant departure from the expected LK theory near $T_N=2.4$~K: both a peak-like anomaly and an enhancement in the temperature dependence of quantum oscillation amplitude are seen in SmSb. Moreover, we discovered a remarkable sensitivity of the SdH amplitudes to sample purity. By adjusting the sample purity, we were able to tune the temperature dependence of the $\alpha$ band's SdH amplitudes from a peak-like anomalous behavior to an enhancement. Therefore, SdH oscillations from the $\alpha$ band connect the two well-known non-LK behaviours, controllable through varying the sample purity, paving the way for developing further understanding of the mechanism leading to the anomalous quantum oscillations.

\end{abstract}

\maketitle

\section{I. Introduction}

Magnetic quantum oscillations such as the de Haas-van Alphen effect (dHvA) and the Shubnikov-de Haas effect (SdH) are important tools to study metallic states. These oscillations, displaying the characteristic periodicity in the inversed magnetic field, enable the determination of the Fermi surface size and shape, the scattering rates, the quasiparticle effective masses ($m^*$) and the Berry phase~\cite{Berry1984,Shoenberg_book,Xiao2010}. Thus, quantum oscillations have become an indispensable tool to study a wide range of exotic metals ranging from heavy fermion systems to surface states of topological insulators~\cite{McCollam2004,Shishido2018,Li2014,Tan2015,Mun2015,Guo2021,Hu2018,Hu2020,Zhao2019,Qu2010,Miyake2006,Aoki1993,Daou2006,Liu2022,Hsu2021}.

Quantum oscillation signals can be weakened by various factors such as high temperatures and sample disorders. In the Lifshitz-Kosevich (LK) theory, the suppression of quantum oscillation amplitudes is described by a product of damping factors. The thermal damping factor, commonly denoted as $R_T$, describes the weakening of the quantum oscillation amplitude as temperature increases. This damping factor is a direct consequence of the smearing of the Fermi-Dirac distribution function near the Fermi energy when thermal energy is injected, and $R_T=X/\sinh X$ where $X=2\pi^2k_Bm^*T/e\hbar B$, $k_B$, $e$ and $\hbar$ are the Boltzmann constant, elementary charge and reduced Planck constant, respectively. $T$ is the temperature and $B$ is the applied magnetic field~\cite{Shoenberg_book}. Hence, $R_T$ is frequently used to extract the quasiparticle effective masses, and has played a prominent role in strongly correlated electron systems where $m^*$ is usually strongly renormalized~\cite{McCollam2004,Shishido2018,Li2014,Tan2015,Mun2015,Guo2021,Hu2018,Hu2020,Zhao2019,Qu2010,Miyake2006,Aoki1993,Daou2006}.

In a majority of cases, the temperature dependence of quantum oscillation amplitudes can be very well described by $R_T$, which is a smooth and monotonically decreasing function of $T$ for a fixed $B$ and $m^*$. However, several important cases that show spectacular deviation from the expected $R_T$ have been reported, giving rise to the so-called non-LK behaviour~\cite{McCollam2004,Shishido2018,Li2014,Tan2015,Hartstein2018}.  
Since an experimental agreement with $R_T$ directly confirms the existence of quasiparticles governed by Fermi-Dirac statistics, a violation of LK theory hints at an unconventional ground state.
Two distinct types of non-LK behaviors have been reported. 
For the first type, the quantum oscillation amplitude is anomalously suppressed at the low-temperature region. Thus, the temperature-dependent amplitude shows a peak, defying the monotonic variation required by $R_T$. We call this a ``peak-like" anomaly. 
Such type of 
non-LK behaviour has been observed in the heavy fermion superconductor CeCoIn$_5$, and the non-LK behaviour was first attributed to spin-dependent masses while a recent treatment argues that the peak is due to the entrance into a field-induced antiferromagnetic (AFM) ground state~\cite{McCollam2004,Shishido2018}.
The second type of non-LK behaviour is an enhancement anomaly, in which low-temperature quantum oscillation amplitudes grows well beyond the expectation of $R_T$. This is most famously observed in SmB$_6$, believed to be a topological Kondo insulator, 
\ie a bulk correlated insulator with a conductive surface.
Although the amplitude can be nicely described by $R_T$ above $\sim$1~K, a dramatic enhancement of oscillation amplitudes sets in below $\sim$1~K, 
without a sign of saturation even as $T\rightarrow$~0~K~\cite{Tan2015}. The origin of the non-LK behaviour in SmB$_6$ is not yet settled. Therefore, the observation has aroused intense theoretical interest~\cite{Knolle2015,Knolle2017,Harrison2018,Pal2016,Pal2017,Pal2019,Peters2019,Zhang2016}.

Recently, intriguing non-LK behaviour has been reported in SmSb~\cite{Wu2019}. SmSb belongs to a large family of semimetals with a NaCl-type structure, and several members of this family have been proposed to be topologically nontrivial semimetals,~\eg Refs.~\cite{Tafti2016,Han2017,Zeng2016,Mullen1974,Nayak2017,Kuroda2018,Fang2020}. 
The LK violation displayed by SmSb~\cite{Wu2019} is unusual in two regards: first, the LK violation is observed only in SdH but not dHvA; second, both the peak and enhancement anomalies are observed in SdH, but for two different frequencies, i.e., the two distinct sets of bands undergoing quantum oscillations lead to two different types of non-LK behaviour.

In this work, we report our measurements on the SdH and the dHvA effects for 7 SmSb crystals, spanning a range of crystal qualities. We discover a surprising interplay between crystal quality and LK violation.
Our data show that, with a decreasing crystal quality, the temperature dependence of the SdH amplitudes associated with the $\alpha$ bands evolve from the peak anomaly observed in Ref.\ \onlinecite{Wu2019} to an enhancement anomaly not reported before. Meanwhile, such a sensitivity to the crystal quality is absent in the other aspects of quantum oscillations, namely, an enhancement anomaly is observed for the $\beta$ bands and the dHvA follows LK well. Our results establish crystal quality as a tunable knob between the two types of non-LK behaviours in the same material for the same electron band, and highlights the prominence role played by disorder in LK violation.\\

\begin{figure}[!t]\centering
      \resizebox{9cm}{!}{
              \includegraphics{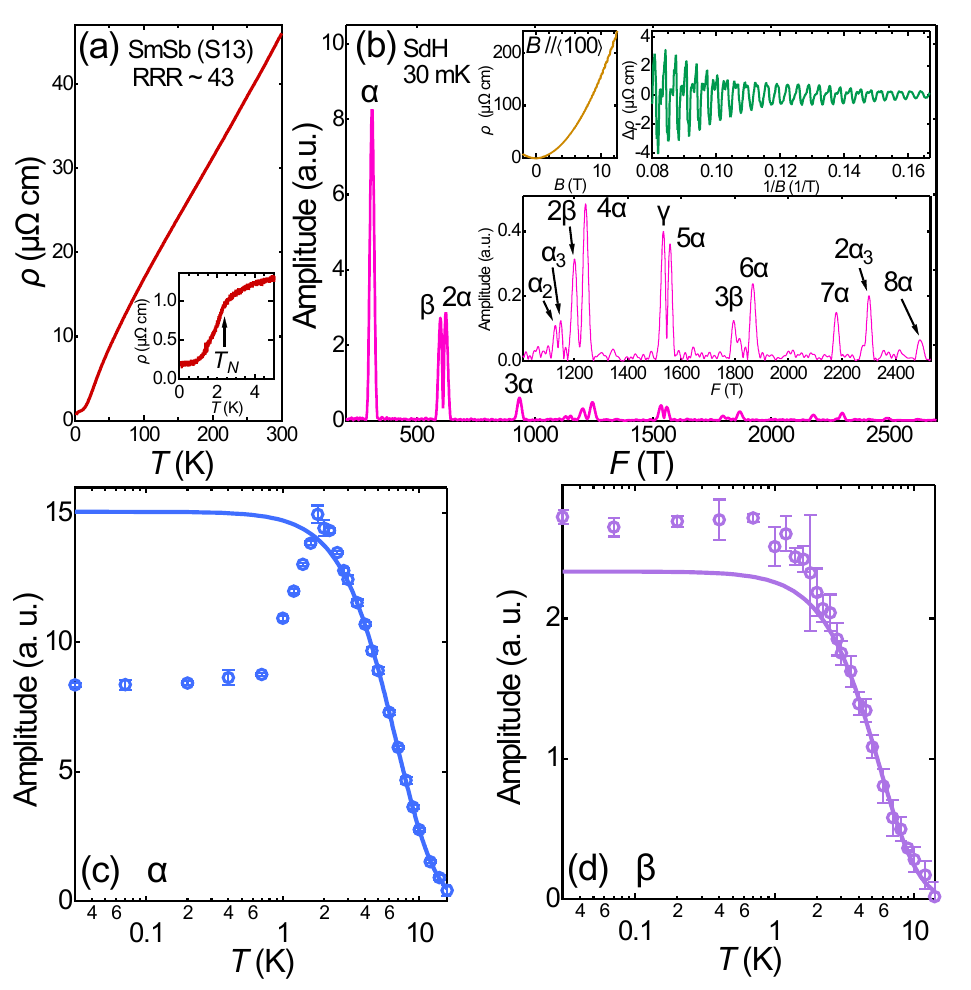}} 			
              \caption{\label{fig1}
              (a) Temperature dependence of resistivity $\rho(T)$ in SmSb\ (S13) and the inset shows the AFM transition with $T_N$ about 2.4 K. (b) FFT spectra of SdH quantum oscillation at 30 mK with field range between 6~T to 12.5~T. The upper insets show the magnetic field dependence of resistivity with $B$ along $\langle100\rangle$ direction and the residual component containing the oscillatory signals. The lower inset is the FFT spectra from 1000~T to 2530~T. Anomalous LK behaviour for (c) $\alpha$ band and (d) $\beta$ band. The solid curves in (c) and (d) are the LK fit above 3~K. The error bars in (c) and (d) are defined as the magnitude of the noise floor in the vicinity of the given peak.}
\end{figure}

\section{II. Methods}

Single crystals of SmSb were grown using Sn flux method. The starting materials of Sm, Sb, and Sn with the molar ratio of 1 : 1 : 20 were put into an alumina crucible and sealed in a silica tube under high vacuum. The reagents were heated to 1323 K, kept at this temperature for 10 h, and slowly cooled to 1073 K over 100 h. The remaining Sn flux was separated by centrifugation and several cubic-shaped crystals with a typical size of $1 \times 1 \times 1~mm^3$ were mechanically removed from crucible. The phase purity and the orientation of crystals were characterized by powder X-ray diffraction (Bruker D2 phaser).

The electrical resistivity was measured by a standard four-terminal configuration in a Physical Property Measurement System by Quantum Design (down to 2 K) or a dilution refrigerator by Bluefors (down to 30 mK). The dHvA measurement was performed by a standard modulation field method with a 800-turn drive coil and a 10-turn pick-up coil. 

 Band structure calculations were performed using the WIEN2k package~\cite{Schwarz2003}, which implements the DFT with all-electron full-potential linearized augmented plane-wave. Experimental lattice constants from Ref.~\cite{ Abulkhaev1992} were adopted in the calculations. The generalized gradient approximation (GGA) of Perdew, Burke, and Ernzerhof (PBE)~\cite{Perdew1996} was employed as the exchange-correlation potential. $R_{MT}^{\min}K_{\max}=9.0$ and a $k$-point mesh of 8000 in the first Brillouin zone were used. The muffin-tin radius for Sm and Sb atoms were set to 2.5 a.u. and the 4$f$ electrons of Sm were treated as core states. Spin-orbit coupling was included by using the second variational procedure~\cite{Macdonald1980}. After obtaining the Fermi surfaces from the DFT calculation, the Supercell K-space External Area Finder (SKEAF) code~\cite{Rourke2012} was used to extract the quantum oscillation frequencies.\\

\section{III. Results}

Figure~\ref{fig1}(a) shows the $T$ dependence of the electrical resistivity in one of our SmSb single crystals (S13) with a residual resistance ratio (RRR, defined as $\rho$(290~K)/$\rho$(5~K)) of $\sim$ 43. On cooling, the system undergoes an antiferromagnetic (AFM) transition from the paramagnetic phase at around 2.4~K with $\rho(T)$ dropping abruptly as displayed in the inset of Fig.~\ref{fig1}(a)~\cite{Hulliger1978}. 
Many rare-earth monopnictides exhibit an extremely large magnetoresistance~\cite{Tafti2016a,Han2017,Hu2018,Wu2019,Kumar2016,Fang2020,Xu2017,Tafti2016}, which aids the observation of pronounced SdH signals. In the magnetic field, the magnetoresistance of SmSb (S13) reaches $\sim142,000\%$ at 30 mK and 12.5 T (see upper inset of Fig.~\ref{fig1}(b)). SdH quantum oscillations can be discerned after removing the background (see upper inset of Fig.~\ref{fig1}(b)). To rule out the influence from the large background, the oscillatory signals are normalized by the field-dependent background~\cite{Wu2019,Shoenberg_book}.
The main panel of Fig.~\ref{fig1}(b) shows the fast Fourier transform (FFT) spectrum with an excellent signal-to-noise ratio for the dataset at 30~mK between 6~T to 12.5~T. The quantum oscillation frequencies for $\alpha$ (311~T), $\beta$ (600~T) and $2\alpha$ (623~T) are consistent with the values reported~\cite{Wu2019}. Besides, we also detect an abundant of higher order harmonics. For example, $8\alpha$ (2490~T) can be seen in SmSb (S13), as displayed in the lower inset of Fig.~\ref{fig1}(b), indicating a high-quality single crystal. The frequencies labelled $\alpha_2$ (1132~T) and $\alpha_3$ (1150~T) comes from the two identical but mutually orthogonal electron ellipsoids (see Fig.~\ref{fig4}(c)). As we will see below, the SdH spectrum is consistent with the calculated Fermi surfaces.

To gain more insight into the non-LK behaviour in SmSb, we measure SdH oscillations at different temperatures. As displayed in Figs.~\ref{fig1}(c) and \ref{fig1}(d), for both $\alpha$ and $\beta$, the temperature dependence of the quantum oscillation amplitudes follow the LK theory well in the paramagnetic (PM) phase, and the cyclotron effective masses ($m^*$) obtained by the LK fit above 3~K are 0.20(1)~$m_e$ and 0.25(1)~$m_e$ for $\alpha$ and $\beta$, respectively. The small cyclotron effective masses align well with the observation in other rare-earth monopnictides~\cite{Wu2019, Hu2018}. However, as the temperature approaches the AFM transition, drastic deviations from the LK behaviour appear -- an obvious enhancement appears in $\beta$ and a significant suppression shows up in $\alpha$, leaving an anomalous peak around 2~K. Therefore, we successfully and independently confirm the reported non-LK behaviour~\cite{Wu2019}, setting the scene for a further exploration of this matter in SmSb. Furthermore, we also successfully trace the temperature dependence of the SdH amplitudes for $\alpha_3$ and $\gamma$ (see Supplementary Note 5). As shown in Supplementary Figure 5, $\alpha_3$ and $\gamma$ also display the peak-like anomaly, which is similar to $\alpha$.

\begin{figure}[!t]\centering
      \resizebox{9cm}{!}{
              \includegraphics{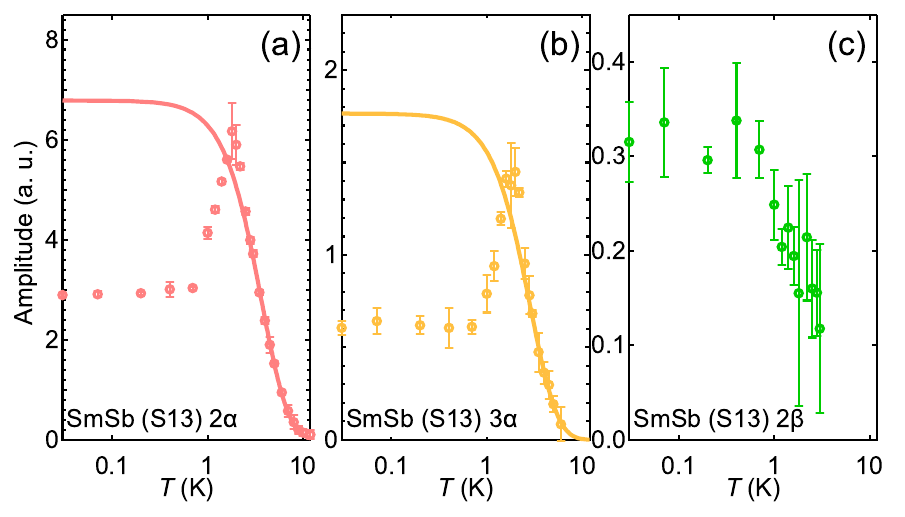}}            			
              \caption{\label{fig2}
             Temperature dependence of SdH quantum oscillation amplitudes in SmSb (S13) for (a) $2\alpha$, (b) $3\alpha$ and (c) $2\beta$. The solid curves in (a) and (b) are the LK fit above 3~K. The error bars are defined as the magnitude of the noise floor in the vicinity of the given peak.
              }            
\end{figure}

With higher order harmonics detected, we next investigate if the non-LK behaviour can also be seen for the harmonics. From the FFT spectrum in Fig.~\ref{fig1}(b), $2\alpha$ has a good signal-to-noise ratio, comparable to that of $\beta$ at 30~mK. Despite the very similar frequency values for $2\alpha$ and $\beta$, the amplitude of $2\alpha$ (Fig.~\ref{fig2}(a)) strongly resembles the temperature dependence seen in $\alpha$ instead of $\beta$. Similarly, the oscillation amplitude of $3\alpha$ also shows a peak at around 2~K. Also shown in Figs.~\ref{fig2}(a) and (b) are the analysis of the data in the PM region (\ie\ above 3~K) with $R_T$, giving the values 0.38(1)~$m_e$ and 0.48(2)~$m_e$ for $2\alpha$ and $3\alpha$, respectively.
Here, $m^*(2\alpha)\approx 2 m^*(\alpha)$, as expected for the effective mass of harmonics. However, $m^*(3\alpha)$ is far below $3 m^*(\alpha)$. This could be due to the much lower signal-to-noise ratio for $3\alpha$, which could affect the accuracy of $m^*$. Unfortunately, a similar effective mass analysis of $2\beta$ cannot be meaningfully carried out due to the weak oscillation signals, and SdH oscillations of $2\beta$ can only be seen up to 3~K (Fig.~\ref{fig2}(c)). Nevertheless, the amplitude of 2$\beta$ does not resemble a peak-like temperature dependence. Finally, based on the overall behaviour of the SdH amplitudes, we conclude that the $R_T$ expression describes the PM region accurately.\\

\begin{figure}[!t]\centering
       \resizebox{8.9cm}{!}{
              \includegraphics{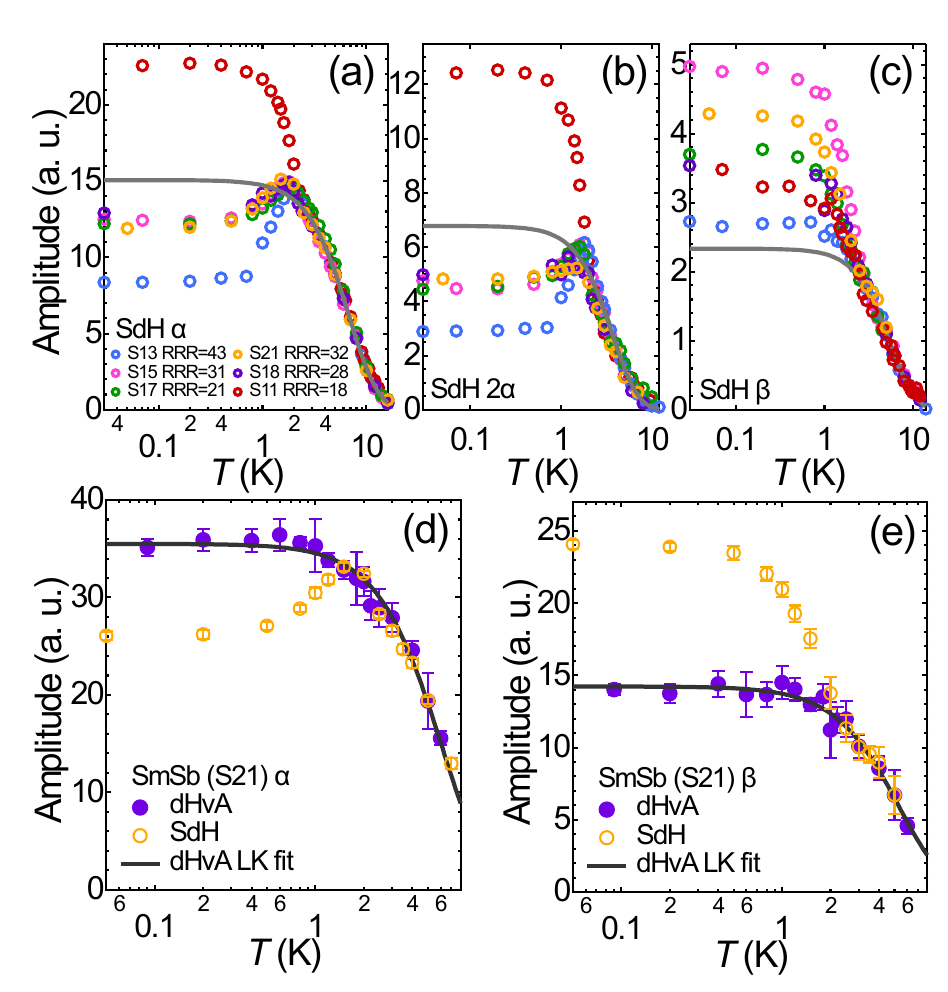}}              				
              \caption{\label{fig3}
              Temperature dependence of SdH quantum oscillation amplitudes for different SmSb samples in (a) $\alpha$, (b) 2$\alpha$ and (c) $\beta$. The amplitudes above 3~K are re-scaled for S15, S21, S18, S17 and S11 to match the amplitudes of S13. The solid grey curves indicate LK fit of SmSb (S13) above 3~K. Temperature dependence of SdH and dHvA quantum oscillation amplitudes of SmSb (S21) in (d) $\alpha$ band and (e) $\beta$ band. The solid black curves indicate LK fit of dHvA from 90~mK to 6~K. In this figure, magnetic field from 6 T to 12.5 T is chosen for the FFT analysis. The error bars in (d) and (e) are defined as the magnitude of the noise floor in the vicinity of the given peak.}
\end{figure}

We now present the central finding of our work. SdH oscillation amplitudes for six samples, spanning RRR values from 18 to 43, are displayed against temperature in Figs.~\ref{fig3}(a) and \ref{fig3}(b) for $\alpha$ and 2$\alpha$, respectively. Data from S13 are also included in these figures. While the FFT spectrum of the sample with the largest RRR (S13) is displayed in Fig.~\ref{fig1}(b), additional FFT spectra for the sample with the lowest RRR (S11) can be found in Supplementary Information (See Supplementary Note 4). The FFT spectra for S11 exhibit an excellent signal-to-noise ratio despite having the lowest RRR. In the PM state, the SdH data can again be accurately described by $R_T$ (solid line), and an excellent agreement can be achieved for all samples. However, striking differences appear in the AFM state. For $\alpha$, the peak-like behaviour progressively changes to an enhancement behaviour when RRR decreases. Meanwhile, the temperature dependence of amplitudes for $2\alpha$ also shows a similar RRR dependence (Fig.~\ref{fig3}(b)). Such an extraordinary sensitivity of the temperature-dependent SdH amplitudes to sample quality is uncommon. In particular, the observed link between the peak-like non-LK behaviour and the enhancement non-LK behaviour, tunable by the sample purity and hosted by the {\it single} SdH frequency, offers an attractive prospect for identifying the underlying mechanism that governs the switching between the two known non-LK behaviours. We note that the antiferromagnetic transition temperature in SmSb is not sensitive to RRR in this RRR range (see Supplementary Note 7). For $\beta$, the enhancement behaviour is observed at low temperatures for all samples, but with very different degrees of amplitude enhancement. Unlike $\alpha$, a systematic trend in the variation of $\beta$ amplitudes has not been observed in this RRR range (Fig.~\ref{fig3}(c)).

\begin{figure}[!t]\centering
       \resizebox{8.8cm}{!}{
    \includegraphics{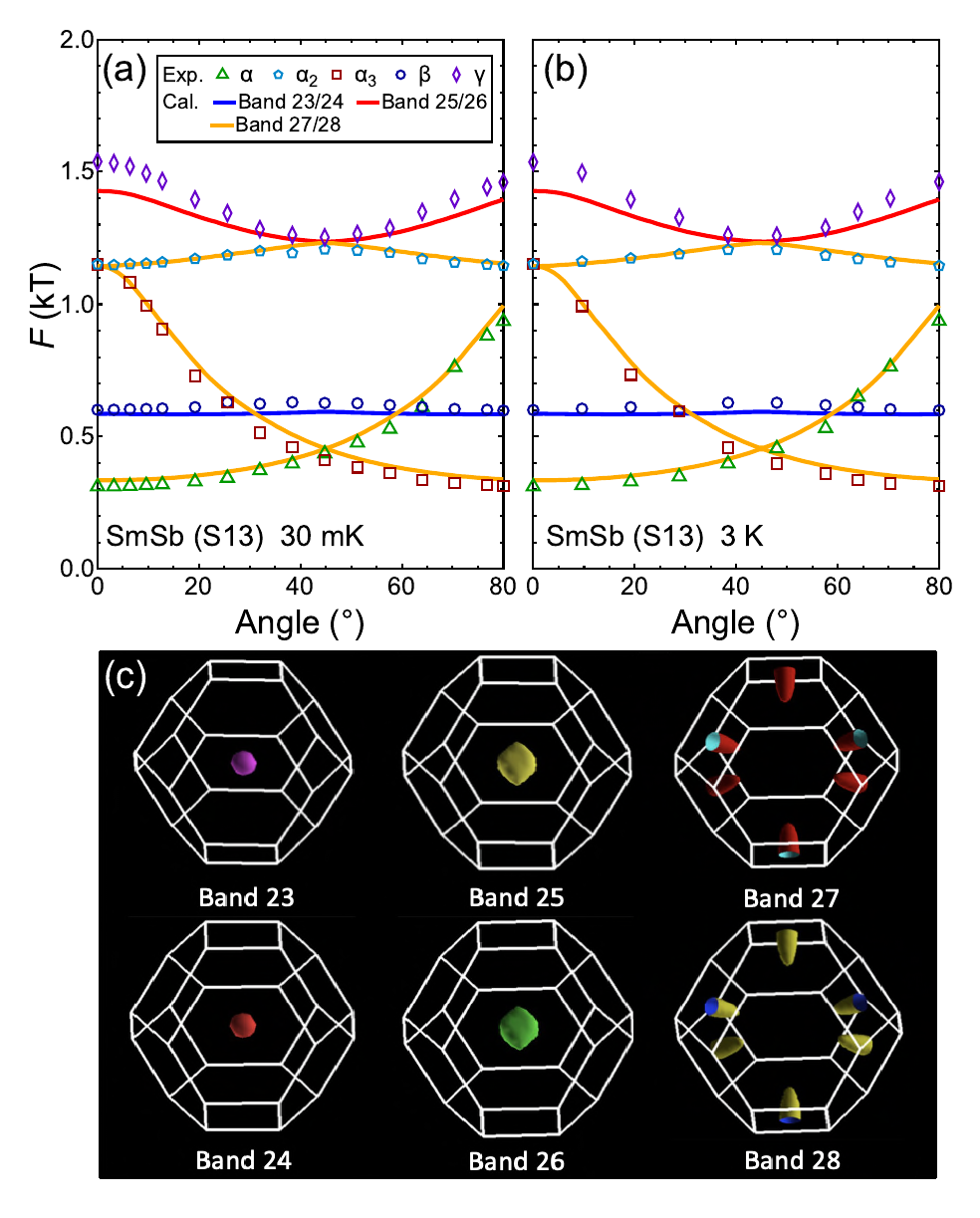}}  \caption{\label{fig4}
              Angular dependence of SdH  oscillation frequencies at (a) 30 mK and (b) 3~K. The solid curves are the calculation results with spin-orbital coupling. (c) Calculated band-resolved Fermi surface sheets of SmSb.}
\end{figure}

Given the observed RRR dependence, one would naturally query if the reported difference between the temperature dependence of the SdH and dHvA amplitudes~\cite{Wu2019} could be attributed to sample dependence. To settle this matter, we conduct both the SdH and dHvA measurements on the {\it same} piece of sample (S21) (Supplementary Note 1). As shown in Figs.~\ref{fig3}(d) and (e), although the non-LK behaviour show up in SdH oscillations for both $\alpha$ and $\beta$, the dHvA results satisfy the LK theory very well. The LK fits down to the lowest temperature give cyclotron effective masses of 0.23(1)~$m_e$ for $\alpha$ and 0.26(1)~$m_e$ for $\beta$, which are consistent with the values determined by high-temperature fit of SdH-determined amplitudes (Figs.~\ref{fig1}(c) and (d)). Finally, we also measure the dHvA effect in another sample (S14) with RRR $\sim$14, and confirm the validity of $R_T$ in describing the amplitudes of both $\alpha$ and $\beta$ (Supplementary Note 2). Our SdH and dHvA results in SmSb rule out sample dependence as the cause for the different behaviour in SdH and dHvA oscillations, and show strong evidence that the non-LK behaviour exist only in SdH oscillations.\\

\section{IV. Discussion}

From our measurements, the deviation from the LK theory in SdH oscillations happens near the AFM transition. We examine if a Fermi surface reconstruction occurs at $T_N$. First, we note that there is no resolvable shift in the SdH frequency across the AFM transition for both $\alpha$ and $\beta$: the numerical upper bound on the change of $\alpha$ and $\beta$ frequencies across $T_N$ are 0.9 T and 2.8 T, respectively, with details presented in Supplementary Note 6. Then, to explore the possible shape change through the AFM transition, we measure the angular dependence of SdH oscillation frequencies $F$ at 30~mK (deep inside the AFM phase) and 3~K (in the PM phase). The experimental data are displayed as symbols in Figs.~\ref{fig4}(a) and (b). Our density functional theory (DFT) calculations, performed for the paramagnetic phase, reveal Fermi surfaces that are characteristic of rare-earth monopnictides~\cite{Hu2018} -- two sets of spin-orbit split pockets centered at the Brillouin zone center and one pair of spin-orbit split ellipsoids at the Brillouin zone boundary. The spin-orbit coupling is negligible, and Bands 23, 25 and 27 are practically identical to their spin-orbit split counterparts Bands 24, 26 and 28, respectively. The quantum oscillation frequencies extracted from the calculated Fermi surfaces (solid curves in Figs.~\ref{fig4}(a) and (b)) show an excellent agreement with the experimental data at 30~mK {\it and} 3~K, except for a minor discrepancy for $\gamma$. Therefore, the bulk Fermi surface are not reconstructed by the onset of AFM, and calculations done in the absence of magnetism can accurately describe the Fermiology of SmSb.\\

While we are not in a position to conclusively settle the issue of the non-LK behaviour in SmSb, we discuss some possibilities. In SmB$_6$, a two-channel model has been adopted to describe the non-LK behaviour of the dHvA results~\cite{Harrison2018}. The two channels are assumed to have identical quantum oscillation frequencies but very different effective masses: $\mathscr{A}(T)=C_1R_{T,1}+C_2R_{T,2}$, in which $C_1$ and $C_2$ are the weights for different channels, and $\mathscr{A}(T)$ is the measured amplitudes. We apply the two-channel model to $\alpha$ band of our SdH results (Supplementary Note 3), which successfully captures both the peak-like and enhancement non-LK features for different samples with different RRR. The analysis gives $m^*_1\sim0.2~m_e$, which is consistent with the effective mass obtained from the high temperature fit, and a larger $m^*_2$ (from 0.6(1)~$m_e$ to 1.6(3)~$m_e$). Importantly, in this two-channel model $C_1$ and $C_2$ must have different (same) signs to describe the peak-like (enhancement) behaviour. This suggests that the phase relationship between the quantum oscillations in the two channels might be influenced by the change in the crystal quality, and this could serve as a phenomenological explanation on the observed disorder tuning between the two types of LK violation. Besides, from the two-channel analysis, channel 2 appears to be more sensitive to disorder. This could provide a path towards the eventual understanding of the non-LK behavior in SmSb, if one could identify a mechanism to manipulate channel 2 effectively.

We further comment that the emergence of two different effective masses in the two-channel model could be a consequence of the hybridization between the itinerant electrons and the well-localized Sm $f$-band.
Such a hybridization has been shown by some authors~\cite{Knolle2015,Peters2019,Zhang2016,Harrison2018,Pal2016,Pal2017} to preserve the quantum oscillation frequencies, which is consistent with our measurements at 3~K and 30~mK (Fig.~\ref{fig4}).

However, it remains a challenge to explain the differences between SdH and dHvA results.
Such difference could be reconciled if surface states emerge in the low-temperature AFM phase for which LK violation is observed, as the surface states would have a diminishing contribution to the dHvA signals (via the magnetic susceptibility) compared with the SdH signals (via the electrical conductivity). Indeed, AFM-induced surface states have recently been discovered in the isostructural compound NdBi~\cite{Schrunk2022}. Yet, even if surface states are responsible for the non-LK behaviours in SmSb, it is unclear why the quantum oscillation frequencies stay unchanged for both the $\alpha$ and $\beta$ bands across the AFM transition.

\section{V. Conclusion}
In summary, we have measured the SdH and dHvA effect in SmSb. Our experiments confirmed the existence of non-Lifshitz-Kosevich temperature dependence of SdH amplitudes in $\alpha$ band and $\beta$ band, as well as in harmonics 2$\alpha$ and 3$\alpha$. Importantly, we unravel a striking sensitivity of the non-LK behaviour to the sample purity: for $\alpha$, the peak-like behaviour progressively changes to an enhancement as RRR decreases and such a sensitivity to the crystal quality is absent in the dHvA study. Therefore, SmSb is an unique compound which displays two well-known types of non-LK behaviours, each has been separately seen in the heavy fermion superconductor CeCoIn$_5$ and the topological Kondo insulator SmB$_6$. The fact that the two non-LK behaviours can be controlled by a tunable parameter, namely the sample quality, offers the prospect of developing further understanding of the mechanism leading to the anomalous quantum oscillations in exotic metals.\\

\section{Acknowledgments}

We acknowledge Qun Niu, Alexandre Pourret, Georg Knebel, Nigel R. Cooper, David Khmelnitskii and Robert Peters for discussions. The work was supported by Research Grants Council of Hong Kong (CUHK 14300418, CUHK 14300419, A-CUHK 402/19), CUHK Direct Grant (4053345, 4053299), City University of Hong Kong (9610438), the Ministry of Science and Technology of Taiwan (MOST-109-2112-M-006-013, MOST-110-2124-M-006-006, MOST-110-2124-M-006-010) and the Croucher Foundation (CF21SC01).\\


\providecommand{\noopsort}[1]{}\providecommand{\singleletter}[1]{#1}%

\end{document}